\documentclass[preprint,authoryear,12pt]{elsarticle}

\usepackage{graphicx}
\usepackage{subfigure}
\usepackage{threeparttable}
\usepackage[linesnumbered,boxed]{algorithm2e}
\usepackage{amsmath}
\usepackage{mathrsfs}
\usepackage{amsfonts}
\usepackage{booktabs}

\journal{European Journal of Operational Research}

\begin{document}
\setlength{\baselineskip}{1.5em}

\begin{frontmatter}

\title{Iterated tabu search for the circular open dimension problem}
\author[HUST,Angers]{Zhanghua Fu\corref{cor1}}
\ead{fuzhanghua1984@163.com}
\author[HUST]{Wenqi Huang }
\ead{wqhuang@mail.hust.edu.cn}
\author[HUST]{Zhipeng L\"u}
\ead{zhipeng.lui@gmail.com}
\cortext[cor1]{Corresponding Author, Tel: +33 07 62 35 58 71}
\address[HUST]{School of Computer Science and Technology,
          Huazhong University of Science and Technology, Wuhan ,430074, China}
\address[Angers]{LERIA, Universit\'{e} d'Angers, 2 Boulevard Lavoisier, 49045 Angers, France}


%
%



%

%



\begin{abstract}
This paper mainly investigates the circular open dimension problem (CODP), which consists of packing a set of circles of known radii into a strip of fixed width and unlimited length without overlapping. The objective is to minimize the length of the strip. An iterated tabu search approach, named ITS, is proposed. ITS starts from a randomly generated solution and attempts to gain improvements by a tabu search procedure. After that, if the obtained solution is not feasible, a perturbation operator is subsequently employed to reconstruct the incumbent solution and an acceptance criterion is implemented to determine whether or not accept the perturbed solution. This process is repeated until a feasible solution has been found or the allowed computation time has been elapsed. Computational experiments based on well-known benchmark instances show that ITS produces quite competitive results with respect to the best known results. For 18 representative CODP instances taken from the literature, ITS succeeds in improving 13 best known results within reasonable time. In addition, for another challenging related variant: the problem of packing arbitrary sized circles into a circular container, ITS also succeeds in improving many best known results. Supplementary experiments are also provided to analyze the influence of the perturbation operator, as well as the acceptance criterion.


\end{abstract}


\end{frontmatter}


\section{Introduction}
Cutting and packing (C\&P) problems are widely encountered in practical applications, such as paper industry (Fraser and George, 1994), wireless communication (Adickes et al., 2002), marine transport (Birgin et al., 2005), aircraft designing (Liu and Li, 2010), material cutting (He et al., 2012), etc. They generally consist of cutting (or packing) a set of small items from (or into) a large object so as to minimize the wasted portion. As well-known NP-hard problems, C\&P problems are extremely challenging to solve exactly and thus heuristics which attempt to obtain approximate solutions within reasonable time have been the most proposed approaches for tackling C\&P problems.

As a representative variant of the C\&P family, the circle packing problem (denoted by CPP) is concerned about how to pack a number of circles of known radii into a larger container without overlapping. The objective is to minimize the container size. According to the size of the circles to pack, CPP can be classified into two categories (Castillo et al., 2008): the arbitrary sized circle packing problem (denoted by ACP) and the uniform sized circle packing problem (denoted by UCP). It should be pointed out that ACP has to deal with both continuous and combinatorial features of the problem, while UCP mainly deals with continuous optimization problem. Due to the extremely large-scale combinatorial solution space of ACP, the approaches proposed for ACP are usually very different from those proposed for UCP.

This paper mainly investigates a variant of ACP: the circular open dimension problem (denoted by CODP), which attempts to pack $N(N=1,2,\cdots)$ arbitrary sized circles into a strip of fixed width and unlimited length without overlapping. More precisely, CODP can be formulated as follows.

Given a strip of fixed width $W$ and unlimited length $L$, as well as $N$ arbitrary sized circles $C_i$ of known radii $r_i(i=1,2,\ldots,N)$. Take the origin of two-dimensional Cartesian coordinate system at the midpoint of the container, and denote the coordinates of the midpoint of $C_i$ by $(x_i,y_i)$. The objective of CODP is to obtain a solution $(X,L)$, where $X$ is a configuration denoted by $(x_1,y_1,\ldots, x_i,y_i,\ldots, x_N,y_N)$, such that

\begin{equation}
\begin{array}{llll}
 Minimize\ L,\ subject \ to:\\
 \\
|x_i|+r_i\le 0.5L \quad \forall \ 1\le{i}\le{N},\\
\\
|y_i|+r_i\le 0.5W \quad \forall \ 1\le{i}\le{N},\\
\\
\sqrt{(x_i-x_j)^2+(y_i-y_j)^2}\ge r_i+r_j \quad \forall \ 1\le{i<j}\le{N}.
\end{array}
\end{equation}

The first two constraints state that each circle should not extend outside the container. The third constraint requires that any pair-wise circles can not overlap with each other. $(X,L)$ is termed a feasible solution if it meets all the constraints.

Furthermore, in order to measure the feasibility of a given solution $(X,L)$, we define a penalty function based on the definition of overlaps as follows. For any solution $(X,L)$, there may exist two kinds of overlaps: overlaps between two circles and overlaps between a circle and a border of the strip. Respectively, the overlapping depth between the $i$th circle $C_i$ and the $j$th circle $C_i$ is

\begin{equation}
O_{ij} = Max\bigg\{0, r_i+r_j-\sqrt{(x_i-x_j)^2+(y_i-y_j)^2} \bigg\}.
\end{equation}

And the overlapping depth between $C_i$ and a vertical border of the strip is
\begin{equation}
O_{ix} = Max\big\{0, r_i+|x_i|-0.5L \big\}.
\end{equation}

Similarly, the overlapping depth between $C_i$ and a horizontal border of the strip is
\begin{equation}
O_{iy} = Max\big\{0, r_i+|y_i|-0.5W \big\}.
\end{equation}

By adding all squares of overlapping depth together, we get a penalty function $E(X,L)$ which measures the feasibility of a solution $(X,L)$ as follows

\begin{equation}
E(X,L)=\sum_{i=1}^{N-1}\sum_{j=i+1}^{N}O_{ij}^2+\sum_{i=1}^{N}(O_{ix}^2+O_{iy}^2).
\end{equation}

According to this definition, it is not difficult to find that: (1) For any solution $(X,L)$, $E(X,L)\ge0$. (2) $(X,L)$ is feasible if and only if $E(X,L)=0$. Therefore, the objective of CODP is to minimize $L$ while guaranteeing $E(X,L)$=0.

In the present work, an iterated tabu search approach named ITS is proposed for tackling CODP. ITS is composed of a tabu search procedure and a perturbation operator associated with an acceptance criterion. Specifically, ITS forcibly squeezes all the circles into the strip at first and then attempts to gain further improvements by repeatedly performing tabu search and solution perturbation, until a feasible solution has been obtained or the allowed computation time is elapsed. Computational results demonstrate that ITS is rather competitive with respect to the state-of-the-art approaches, in terms of both solution quality and computation time.

The rest of this paper is organized as follows. Section 2 briefly reviews the relevant literature. Section 3 presents the details of the proposed ITS algorithm. Computational experiments and analysis are presented in Section 4 and Section 5 concludes the paper.

\section{Related literature}
Various approaches have been developed for solving the circle packing problems (CPP). In this section, we briefly review the approaches proposed for solving CPP, especially for the circular open dimension problem (CODP) and its closely related variants.

As mentioned above, CPP can be classified into two categories: the arbitrary sized circle packing problem (ACP) and the uniform sized circle packing problem (UCP). Herein, we introduce several representative approaches proposed for ACP at first and then briefly review the algorithms for UCP subsequently.

Almost all the competitive approaches for solving ACP are heuristics, which can be mainly classified into two categories: constructive approaches and perturbation-based approaches. Constructive approaches attempt to pack the circles one by one in sequence into the container according to some constructive rules, until all the circles have been packed feasibly. In contrast, perturbation-based approaches start from one or several initial configuration(s) which contain(s) all the circles (generally with overlapping) and then attempt to gain further improvements by continuous optimization and solution perturbations, until a feasible solution has been obtained. Obviously, these two kinds of strategies are essentially different from each other.

As we know, most of the existing approaches for ACP are constructive approaches. Respectively, for the version of strip or rectangular container (CODP and its variant), many constructive approaches have been proposed. For example, George et al. (1995) developed several heuristic building rules to simulate the packing process, including a quasi-random technique and a genetic algorithm. Hifi and M'Hallah (2004) proposed a constructive procedure and a genetic algorithm. Huang et al. (2005) developed two greedy approaches denoted by B1.0 and B1.5 based on the \emph{maximum hole degree} rule. In order to improve B1.0 and B1.5, Kubach et al. (2009) developed several greedy algorithms and parallelized them by a master slave approach followed by a subtree-distribution model. Kallrath (2009) also applied several heuristics, including a branch and reduce optimization navigator, a column enumeration approach and a symmetry constraints breaking strategy. Akeb and Hifi (2008) proposed an open strip generation solution, an exchange-order strategy to augment the first heuristic and a hybrid heuristic that combines beam search with a series of predetermined interval search. For further improvements, Akeb et al. (2011) proposed an augmented algorithm which incorporates a beam search, a binary search, a multi-start strategy and a separate-beams strategy. And they also proposed an adaptive look-ahead strategy-based algorithm (Akeb and Hifi, 2010). Moreover, for the situation of circular container, there are also several constructive approaches, such as the corner occupying algorithm A1.0 and A1.5 (Huang et al., 2006), the beam-search strategy BS (Akeb et al., 2009), the adaptive beam search look-ahead algorithm (Akeb et al., 2010), the adaptive hybrid algorithm TS/NP (Al-Modahka et al., 2011), etc.

Meanwhile, there are various perturbation-based algorithms proposed for solving ACP. Respectively, for the version of strip or rectangular container, Stoyan and Yaskov (1998; 2004) used the reduced gradient method for local optimization and developed several strategies for transition from one local minimum to another one. Hifi et al. (2004) proposed a simulated annealing approach which also combines the gradient descent method with several configuration transformation strategies. Moreover, for the situation of circular container, several efficient perturbation-based algorithms exist, such as the quasi-physical quasi-human algorithm (Wang et al., 2002), the population basin hopping method (Addis et al., 2008a), the simulated annealing approach (M\"uller et al., 2009), and the energy landscape paving method (Liu et al., 2009), etc.

On the other hand, UCP has also been extensively investigated and many efficient approaches exist, such as the non-linear programming solver(MINOS) (Maranas et al., 1995), the billiard simulation approach (Boll et al., 2000), the population basin hopping method (Addis et al., 2008b; Grosso et al., 2010), the greedy vacancy search strategy (Huang and Ye, 2010), the quasi-physical global optimization method (Huang and Ye, 2011), etc. In addition, as mentioned above, due to the extremely challenging combinatorial feature of ACP, the approaches proposed for ACP are usually quite different from those proposed for UCP.

Finally, we refer interested readers to (W\"ascher et al., 2007; Castillo et al., 2008;  Hifi and M'Hallah, 2009) for more comprehensive reviews about the C\&P problems.

\section{Proposed approach}

This paper mainly studies CODP, which considers how to pack a number of arbitrary sized circles into a strip of fixed width and unlimited length without overlapping, so as to minimize the length of the container. As a representative variant of ACP, CODP should deal with both the continuous and combinatorial features of the problem. Therefore, different continuous or combinatorial optimization strategies usually lead to different approaches for solving CODP.

In this paper, a hybrid meta-heuristic algorithm named ITS is proposed, which integrates a tabu search procedure (TS) and a perturbation operator associated with an acceptance criterion. Respectively, TS is a robust neighborhood search approach, while the perturbation operator associated with the acceptance criterion is employed to drive the search out of local optimum trap towards new promising region of the solution space. In this paper, we present an efficient algorithm with novel combination of these various strategies, as would outlined in  Algorithm 1.

Respectively, the key components of the proposed approach are detailed in the following subsections.

\begin{algorithm}[H]
\label{ITS} \caption{Outline of the proposed approach}
\KwIn{ Radii of all the circles and the width $W$ of the strip}
\KwOut{Feasible solution $(X^{ITS},L^{ITS})$ with as small $L^{ITS}$ as possible}
$L \leftarrow PreSetL()$; \% pre-set the length of the strip, see 3.4\\

\Repeat
{
    $(X,L)$ is feasible or the limited time has been elapsed
}{
\% randomly generate an initial solution \\
$(X,L) \leftarrow RandomInit()$;        \\
\ \% further optimize $(X,L)$ by TS, see 3.1
$(X,L) \leftarrow TabuSearch(X,L)$;        \\
\Repeat
{
$(X,L)$ is feasible or it is not improved for 10 perturbations
}
{
\ \% reconstruct $(X,L)$, see 3.2 \\
$(X^{'},L) \leftarrow Perturb(X,L)$;   \\
\% further optimize $(X^{'},L)$ by TS, see 3.1
$(X^{*},L) \leftarrow TabuSearch(X^{'},L)$;  \\
\% accept $(X^{*},L)$ if and only if it is better than $(X,L)$, see 3.3\\
 \If{$(X^{*},L)\ is\ better\ than\ (X,L)$}
 {
    $(X,L) \leftarrow (X^{*},L)$
 }
}
}
\% post-process $(X,L)$ for further optimization, see 3.4 \\
$(X^{ITS},L^{ITS})\leftarrow PostProcess(X,L)$; \\
\Return $(X^{ITS},L^{ITS})$
\end{algorithm}

\subsection{Tabu search procedure}
Tabu search (TS) is a well-known meta-heuristic (Glover, 1989; Glover, 1990) which has proven to be effective for solving a large number of practical optimization problems, including quadratic assignment (James et al., 2009), unconstrained global optimization (Duarte et al., 2011), course timetabling (L\"u and Hao, 2010), graph coloring (Wu and Hao, 2012), etc. For CODP, this paper employs a TS procedure as follows.

From an initial solution $(X,L)$, TS calls the well-known unconstrained minimization algorithm LBFGS (Liu and Nocedal, 1989) for continuous optimization at first (the role the LBFGS algorithm plays is illustrated in Figure 1), and then attempts to improve $(X,L)$ consistently by iteratively updating $(X,L)$ with its best neighboring solution, with the aid of forbidden rule and aspiration criterion, until the incumbent solution $(X,L)$ is feasible or $(X,L)$ cannot be further improved within a given number of consecutive iterations. For the sake of efficiency, $L$ is pre-set to a proper constant which remains unchanged until the search process terminates, then, some post-process techniques is used to further optimize $L$ as much as possible (as detailed in subsection 3.4).

\begin{figure}
\begin{center}
\includegraphics[width=3in]{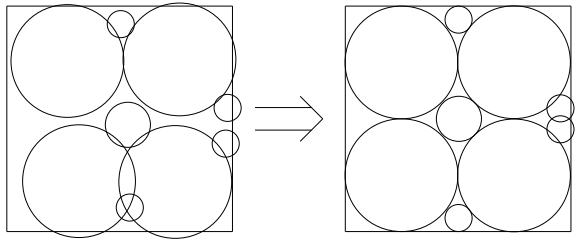}
\caption{Continuous optimization by LBFGS}
\end{center}\vspace{-4mm}
\end{figure}

Like other neighborhood search-based approaches, one of the most important features of TS is the definition of its neighborhood. For CODP, we note that swapping two circles with different radii usually leads to a new candidate solution. Therefore, we introduce a constrained neighborhood $N(X,L)$ of $(X,L)$ as follows.

$N(X,L)$ is a solution set which contains all the neighboring solutions of $(X,L)$, where a neighboring solution is generated by swapping a pair-wise circles $C_i$ and $C_j$ with similar radii and calling LBFGS subsequently for continuous optimization (denoted by $LBFGS((X,L)\oplus Swap(C_i,C_j))$). More precisely, if all the circles are sorted in descending order according to their radii, $C_i$ and $C_j$ are with similar radii means $|j-i|\le2$ and $r_i\neq r_j$.

$N(X,L)$ is formally identified as

\begin{equation}
N(X,L)=\big\{LBFGS((X,L)\oplus Swap(C_i,C_j)), \forall \ |j-i|\le2 \ and \  r_i\neq r_j \big \}.
\end{equation}

On one hand, the circles with similar radii usually play similar roles in the entire solution so that swapping them may obtain further improvements without destroying the entire solution too much. On the other hand, although a larger neighborhood usually leads to a better solution, but more computational efforts are also needed. In consideration of both solution quality and computing complexity, we generate neighboring solutions by swapping two circles with similar radii, instead of swapping any pair-wise circles.

Furthermore, it is usually necessary for a neighborhood search-based approach to prevent local cycling during the recursive neighborhood search iterations. Therefore, a special data structure $tabu \ list$ is introduced to forbid the previously swapped circles to be re-swapped within a certain number of iterations. Technically, the $tabu\ list$ is presented by an integer array $TabuTenure[N]$ (initialized to be zero) which records the tabu tenures of all the circles. $TabuTenure[i]<CurIterNum$ (the current iteration) means $C_i$ is a free circle which can be swapped freely to generate neighboring solutions at the current iteration. Otherwise, $C_i$ is a tabu circle which is forbidden to be swapped until the $TabuTenure[i]$ iteration.

At each iteration of the TS procedure, if the best neighboring solution corresponds to swapping $C_i$ and $C_j$, then the tabu tenures of $C_i$ and $C_j$ should be updated respectively as follows

\begin{equation}
\begin{array}{ll}
 TabuTenure[i] = CurIterNum+T+rand(\frac{N}{8}),\\
 \\
 TabuTenure[j] = CurIterNum+T+rand(\frac{N}{8}).
\end{array}
\end{equation}

It means that $C_i$ (respectively, $C_j$) will be forbidden to be swapped with any other circle for $T+rand(\frac{N}{8})$ iterations, where $T$ is a constant which is experimentally fixed at 2 and $rand(\frac{N}{8})$ denotes a randomly generated integer from 0 to $\frac{N}{8}$ (rounded to the nearest integer).

With this forbidden rule, the TS procedure then restricts consideration to free circles which are not forbidden by the $tabu\ list$. It updates the incumbent solution with its best neighboring solution iteratively. However, in some occasions, swapping some tabu circles may lead to an excellent solution which has not been visited. In this case, it is necessary to readmit this solution as an element of the neighborhood. In order to mitigate this problem, such an aspiration criterion is further employed to override the forbidden rule: if a swapping leads to a solution better than (with lower penalty function) the best solution found so far, the generated neighboring solution will be admitted as a candidate solution, no matter the swapped circles are tabu or not.

The set which contains all the neighboring solutions excluded by the forbidden rule is denoted by $F(X,L)$, and the set which contains all the neighboring solutions readmitted by the aspiration criterion is denoted by $A(X,L)$. Then, the final neighborhood is redefined as $N_{FA}(X,L)$, which is a subset of the previously defined neighborhood $N(X,L)$.

\begin{equation}
N_{FA}(X,L)=\big(N(X,L)-F(X,L)\big)\bigcup A(X,L).
\end{equation}

\begin{algorithm}[H]
\caption{Tabu search procedure for solving CODP}
\KwIn{The incumbent solution $(X,L)$}
\KwOut{Best solution encountered so far}
\For{i=1\ To\ N}
{
    $TabuTenure[i]=0;$ \quad \% initialize the tabu tenures\\
}

\Repeat
{
    $(X,L)$ is feasible or it has not been improved for 20 iterations
}{
   \% generate neighborhood of $(X,L)$\\
    $N_{FA}(X,L)\leftarrow GenerateNeighborhood(X,L);$  \\
     \% update $(X,L)$ with its best neighboring solution\\
    $(X,L) \leftarrow UpdateWithBestSolution(N_{FA}(X,L));$ \\
    \% update the tabu tenures of the corresponding circles\\
    $UpdateTabuTenure(TabuTenure);$ \\
}
 \Return $(X,L)$;
\end{algorithm}

With the definition of $N_{FA}(X,L)$, TS iteratively updates the incumbent solution $(X,L)$ with the best solution (with lowest penalty function) of its neighborhood $N_{FA}(X,L)$, until a feasible solution $(X,L)$ with $E(X,L)=0$ has been obtained, or $(X,L)$ cannot be further improved for a certain number of consecutive iterations (e.g. 20 iterations). The framework of TS is described in Algorithm 2.

\subsection{Perturbation operator}
Like other neighborhood search-based strategies, TS realizes the intensification search which optimizes the objective function as far as possible within a limited search region. However, it usually falls into local optimum trap even with the aid of $tabu\ list$. For the sake of diversification, it is preferred to combine TS with some diversification operators that drive the search to explore new promising region of the solution space.

An easily implemented method is to destruct the incumbent solution completely randomly. However, this naive strategy is not efficient enough because it cannot guide the search to move towards new promising solution space based on the incumbent solution. In order to overcome its weakness, a solution perturbation operator is employed to reconstruct the incumbent solution $(X,L)$ strategically as follows:

(1)	Sort all the circles in descending order according to their radii and then categorize each circle as either a large circle or a small circle. For each circle $C_i$, it is termed a large circle if $r_i>\frac{1}{2}r_{avg}$,where $r_{avg}=\frac{\sum_{i=1}^{N}r_i}{N}$ is the average radius of all the circles to pack. Otherwise, $C_i$ is termed a small circle.

(2)	Remove all the small circles from $(X,L)$.

(3)	Randomly swap $SwapNum$ pairs of large circles with similar radii and call LBFGS subsequently for continuous optimization (see subsection 3.1 for the definition of similar radii).

(4)	Pack the removed small circles back into the proper position one by one in a descending order according to their radii. Respectively, for each small circle $C_i$, we randomly replace it into the strip $N$ (the instance size) times and call LBFGS algorithm subsequently to obtain $N$ local optimal solutions. After that, we only retain the best solution (with lowest penalty function), within which $C_i$ is considered to be packed into the proper position.

After the above 4 perturbation steps, $(X,L)$ is reconstructed to a new solution $(X^{'},L)$. Note that $SwapNum$ should be set prudently, due to the fact that swapping too many circles does not perform differently from random restarting, while swapping too few circles usually leads to local cycling. In order to exploit the tradeoff between intensification and diversification, we empirically set $SwapNum$ to be about $\frac{N}{3}$ (rounded to the nearest integer).

As expected, this perturbation operator is able to relocate the small circles into the proper position rapidly, while escaping far enough away from the current local optimum trap. The rationale behind is due to the experience that the structure of a given solution mainly depends on the large circles, while the small circles are usually suitable to be located in some vacant region. We utilize this heuristic to speed up the search procedure.

\subsection{Acceptance criterion}
Moreover, it is important to introduce a robust acceptance criterion which determines whether to accept the perturbed solution or not . This paper designs a concise acceptance criterion as follows: each time after the incumbent solution $(X,L)$ is perturbed to $(X^{'},L)$, it would be further optimized to $(X^*,L)$ by launching TS subsequently. After that, if $E(X^*,L)<E(X,L)$, $(X^*,L)$ is accepted as the new incumbent solution, otherwise, return to $(X,L)$ and repeat a new round of perturbation followed by TS again.

With this acceptance criterion, it is easy to find out that the incumbent solution $(X,L)$ will be improved consistently, due to the fact that only improved solutions may be accepted. We believe that this feature can reinforce the robustness of the proposed approach.

Specifically, in order to guarantee the diversification feature of the proposed approach, once the incumbent solutions has not been further improved for certain (e.g. 10) consecutive rounds of solution perturbations, a multi-start technique is employed to destruct the incumbent solution to a randomly generated initial solution. After that, new rounds of TS followed by solution perturbations are launched again. This process is repeated until a feasible solution has been obtained, or the allowed computation time (e.g. 30 hours) has been elapsed.

\subsection{Pre-setting and post-processing}
So far, we have not discussed how to set the length $L$ of the strip (the width $W$ is fixed). In fact, the value of $L$ should be set carefully due to the fact that if $L$ is too large, the obtained solution is not high-quality enough even if it satisfies all the constraints. Otherwise, if $L$ is too small, it is impossible for any algorithm to obtain a feasible solution such that a lot of computational efforts may be wasted.

In order to determine $L$ properly, this paper develops some pre-setting and post-processing techniques as follows:

(1) Set $L$ to $L^{best}$ , where $L^{best}$ is the best known result reported in the literature.

(2) Launch ITS to return a solution $(X,L)$ with as low penalty energy as possible.

(3) Based on the solution $(X,L)$ returned by ITS, employ a post-processing procedure to optimize $L$ as much as possible, while guaranteeing the feasibility of the obtained solution. Specifically, the post-processing procedure is detailed as Algorithm 3.

\begin{algorithm}[H]
\label{Post-Process} \caption{Post-processing procedure}
\KwIn{Solution $(X,L)$ returned by ITS }
\KwOut{Feasible solution$(X^{ITS},L^{ITS}$) with as small $L^{ITS}$ as possible}
$L^{upper}\leftarrow L+C;$ \quad  \% set the upper bound of L\\
$L^{lower}\leftarrow L-C;$ \quad  \% set the lower bound of L\\
$X^{ITS}\leftarrow X;$ \\

\% optimize $L^{ITS}$ in dichotomous way \\
\Repeat{$L^{upper}-L^{lower} <10^{-4}$}
{
    $L^{ITS}\leftarrow \frac{L^{upper}+L^{lower}}{2};$ \\
    \% call TS for further improvements, see 3.1 \\
    $(X^{ITS},L^{ITS})\leftarrow TabuSearch(X^{ITS},L^{ITS});$   \\
    \eIf{$E(X^{ITS},L^{ITS})=0$}
    {
        $L^{upper}\leftarrow L^{ITS}$\;
    }
    {
        $L^{lower}\leftarrow L^{ITS}$\;
    }
}

\% guarantee that no overlap exists within the final solution\\
\Repeat
{
$E(X^{ITS},L^{ITS})=0$
}
{
   $L^{ITS}\leftarrow L^{ITS}+10^{-5};$ \\
    \% call LBFGS for continuous optimization, see 3.1 \\
    $(X^{ITS},L^{ITS})\leftarrow LBFGS(X^{ITS},L^{ITS})$;  \\
}
\Return $(X^{ITS},L^{ITS})$;
\end{algorithm}

In Algorithm 3, $C$ is a constant which should be large enough to make sure that it is very easy to obtain a strictly feasible solution if $L^{ITS}\leftarrow L^{upper}$, and it is very difficult to obtain a strictly feasible solution if $L^{ITS}\leftarrow L^{lower}$. In this paper, $C$ is uniformly set to $1$. Algorithm 3 returns a strictly feasible solution $(X^{ITS},L^{ITS})$ with as small strip length $L^{ITS}$ as possible. Uniformly, $L^{ITS}$ is rounded to keep 4 significant digits after the decimal point, e.g., $L^{ITS}= \sqrt{2}=1.41421356\ldots$ is rounded to 1.4143 instead of 1.4142 in order to guarantee the feasibility of the obtained solution.

Based on the above-mentioned strategies, the tabu search procedure is augmented to an iterated tabu search algorithm named ITS, as shown in Algorithm 1. Furthermore, we would like to emphasize several key points of the proposed approach. As described above, ITS is composed of a tabu search procedure and a solution perturbation operator, associated with an acceptance criterion. In our opinions, the tabu search procedure should be robust enough to ensure that the incumbent solution can be improved consistently. On the other hand, for the sake of diversification, the perturbation operator associated with the acceptance criterion should be well-designed in order to drive the search towards new promising solution space. In a word, an effective perturbation-based approach for tackling CODP usually corresponds to a rational tradeoff between intensification and diversification.

\section{Results and analysis}
In order to evaluate the performance of the proposed approach, we implement ITS in C++ language and run it on a computer with 2.87 GHz CPU and 512 MB RAM.

Two sets of well-known instances, which have been widely used as benchmarks by previous approaches, are taken from the literature and tested. The first set of 6 instances named SY1-SY6 are taken from Stoyan and Yaskov (1998), and the second set of 12 instances identified as SY12, SY13, SY14, SY23, SY24, SY34, SY56, SY123, SY124, SY134, SY234, SY1234 are taken from Akeb and Hifi (2008). Note that the second set of instances are obtained by concatenating the six original instances of the first set, e.g., SY12 is obtained by concatenating SY1 and SY2, and SY1234 is obtained by concatenating SY1, SY2, SY3, SY4, etc.

\begin{table}
\noindent {\small Table 1\quad Results obtained by ITS with respect to previous approaches}\vspace{-6mm}\\
{\footnotesize
\begin{center} \doublerulesep 0.4pt \tabcolsep 4pt
\begin{tabular}{@{}cccccccccc@{}} \hline \hline
\multicolumn{1}{c}{} &\multicolumn{1}{c}{} &\multicolumn{1}{c}{} &\multicolumn{1}{c}{Others} &\multicolumn{2}{c}{MSBS \& SEP-MSBS} &\multicolumn{2}{c}{A-SEP-MSBS} &\multicolumn{2}{c}{ITS}\\ \cmidrule(r){4-4} \cmidrule(r){5-6} \cmidrule(r){7-8} \cmidrule(r){9-10}
Instance& $N$ &$W$ &$L$ & $L$ &$t_{total}$ &$L$ &$t_{total}$ &$L^{ITS}$ &$t_{total}$ \\ \hline
$SY1$    &30   &9.5   &17.2315           &17.2070          &15600s    &17.0954         &30h        &\textbf{17.0782}  &15115s \\
$SY2$    &20   &8.5   &14.5350           &14.4867          &3510s     &14.4548         &30h        &\textbf{14.4541}  &8538s  \\
$SY3$    &25   &9.0   &14.4670           &14.4176          &8870s     &14.4017         &30h        &\textbf{14.3864}  &8842s  \\
$SY4$    &35   &11.0  &23.5550           &23.4921          &29290s    &\textbf{23.3538}&30h        &23.4177           &30h     \\
$SY5$    &100  &15.0  &\textbf{35.8590}  &36.1818          &30h       &36.0061         &30h        &35.9843           &30h     \\
$SY6$    &100  &19.0  &\textbf{36.4520}  &36.7197          &30h       &36.6629         &30h        &36.6515           &30h     \\
$SY12$   &50   &9.5   &29.7011           &29.6837          &30h       &29.8148         &30h        &\textbf{29.5835}  &18781s \\
$SY13$   &55   &9.5   &30.6371           &30.3705          &30h       &30.4547         &30h        &\textbf{30.3621}  &79168s \\
$SY14$   &65   &11.0  &38.0922           &37.8518          &30h       &\textbf{37.7244}&30h        &37.7512           &30h     \\
$SY23$   &45   &9.0   &27.8708           &\textbf{27.6351} &68460s    &27.7574         &30h        &27.6830           &30h     \\
$SY24$   &55   &11.0  &34.5476           &34.1455          &30h       &34.1511         &30h        &\textbf{34.0701}  &29576s \\
$SY34$   &60   &11.0  &34.9011           &34.6376          &30h       &34.6744         &30h        &\textbf{34.6263}  &15243s \\
$SY56$   &200  &19.0  &64.7246           &65.2012          &30h       &64.7876         &30h        &\textbf{64.5216}  &49126s \\
$SY123$  &75   &9.5   &43.2558           &42.9931          &30h       &43.0930         &30h        &\textbf{42.9566}  &17684s \\
$SY124$  &85   &11.0  &48.8927           &48.8411          &30h       &48.6101         &30h        &\textbf{48.5622}  &46715s \\
$SY134$  &90   &11.0  &49.3954           &49.3254          &30h       &49.2739         &30h        &\textbf{49.2224}  &5259s  \\
$SY234$  &80   &11.0  &45.8880           &45.5576          &30h       &45.4586         &30h        &\textbf{45.4155}  &33586s \\
$SY1234$ &110  &11.0  &60.2613           &60.0564          &30h       &60.3346         &30h        &\textbf{59.9709}  &29595s \\
$Average$ &-   &-     &36.1260           &36.0447          &$\approx$24h     &36.0061  &30h        &\textbf{35.9277}  &$\approx$14h \\
\hline \hline
\end{tabular}
\end{center}} \vspace{2mm}
\end{table}

For each instance, we pre-set $L$ to the best known result $L^{best}$ reported by previous approaches at first, and then attempt to obtain a solution with as low penalty function as possible by launching the proposed ITS algorithm. The search process terminates until any one of the following two stop criterions is met: (1) Successfully terminates if a feasible solution $(X,L)$ with $E(X,L)=0$ is obtained, it means the obtained solution $(X,L)$ is no worse than the best known solution. (2) Abortively terminates if the allowed computation time has elapsed, for an accurate comparison, the limited cumulative computation time is fixed to 30 hours for each instance, just as same as the referenced algorithms. After that, no matter the search process is successfully or abortively terminated, the post-processing procedure described in subsection 3.4 is executed subsequently to optimize $L$ as much as possible while guaranteeing that no overlap exists within the obtained solution, and then the final value of strip length as well as the corresponding configuration is reported as the final solution $(X^{ITS},L^{ITS})$ obtained by ITS.

Table 1 reports the final results $L^{ITS}$ obtained by ITS, with respect to several other state-of-the-art approaches. Respectively, columns 1-2 indicate the name of the instance and its size. Column 3 indicates the width of the strip ($W$) of each instance. Columns 4-8 report the best known results reported by previous approaches, as well as the elapsed computation time. i.e., column 4 indicates the best value of $L$ obtained by either B1.0 \& B1.5 (Huang et al., 2005), or B1.6\_SPP (Kubach et al., 2009), or BSBIS (Akeb and Hifi, 2008). Columns 5-6 indicate the value of $L$ and the cumulative computation time reported by MSBS \& SEP-MSBS (Akeb et al., 2011). Columns 7-8 indicate the value of $L$ and the computation time reported by A-SEP-MSBS (Akeb and Hifi, 2010). For comparison, columns 9-10 indicate the value of $L$ obtained by the proposed ITS algorithm, as well as the cumulative computation time. Note that the results in bold indicate the best ones of all the results obtained by various approaches, and \emph{30h} denotes that the limited 30 hours is elapsed before the corresponding solution was obtained.

As shown in Table 1, ITS succeeds in improving the best known results in 13 occasions out of all the 18 instances, it only fails to match 5 best known solutions after the limited 30 hours is elapsed. With respect to previous approaches, ITS respectively succeeds in improving 16 results reported by A-SEP-MSBS (only except SY4, SY14), 17 by MSBS \& SEP-MSBS (only except SY23), and 16 by the other approaches (only except SY5, SY6). Overall, the average strip length obtained by ITS is 35.9277, which is about 0.22\% better than the average strip length reported by A-SEP-MSBS, about 0.32\% better than MSBS \& SEP-MSBS, about 0.55\% better than the other approaches, respectively. The comparison in term of solution quality undoubtedly indicates that ITS produces competitive results with respect to previous approaches. Figure 2 illustrates the 13 improved solutions (all circles have been sorted in descending order according to their radii), interested readers please contact the authors for the detailed coordinates of each instance.

\begin{figure}
\begin{center}
\includegraphics[width=2.5in]{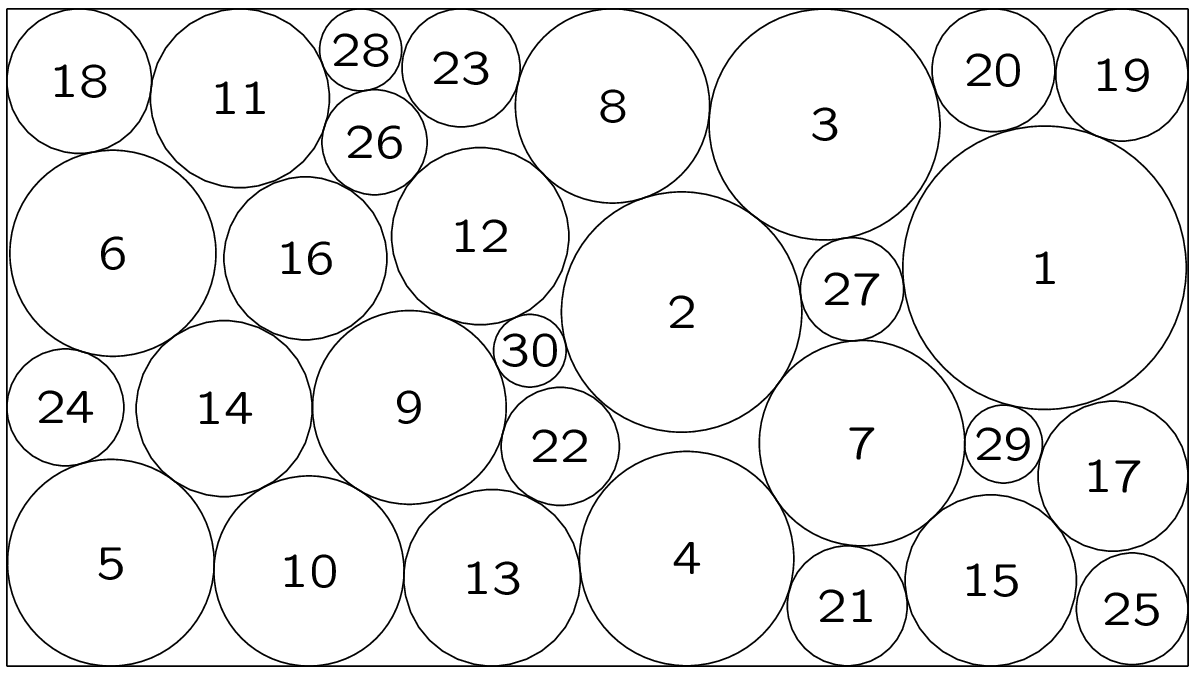}
\includegraphics[width=2.5in]{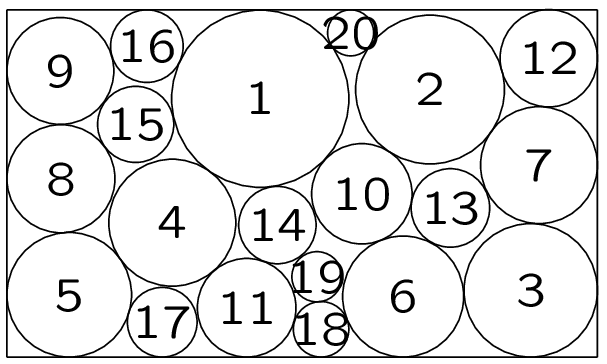}\\
\includegraphics[width=2.5in]{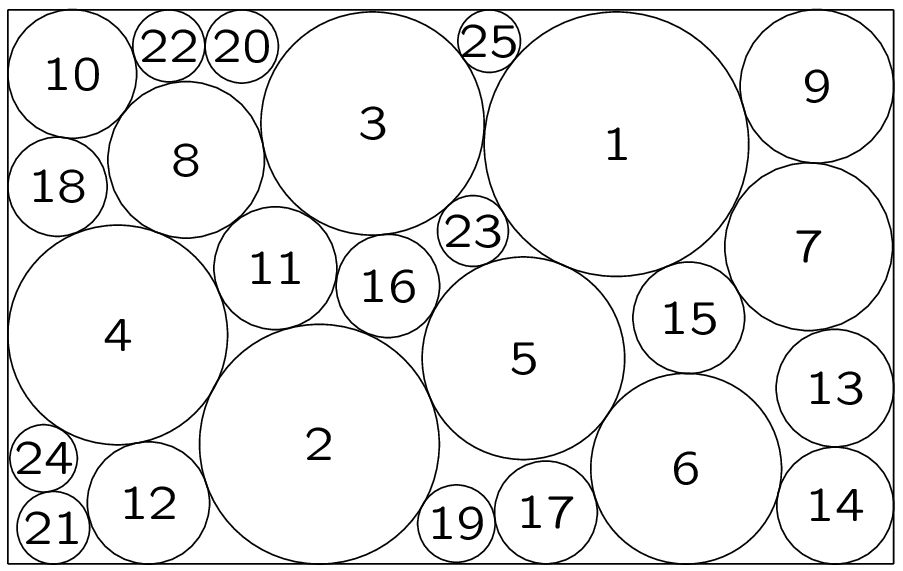}
\includegraphics[width=2.5in]{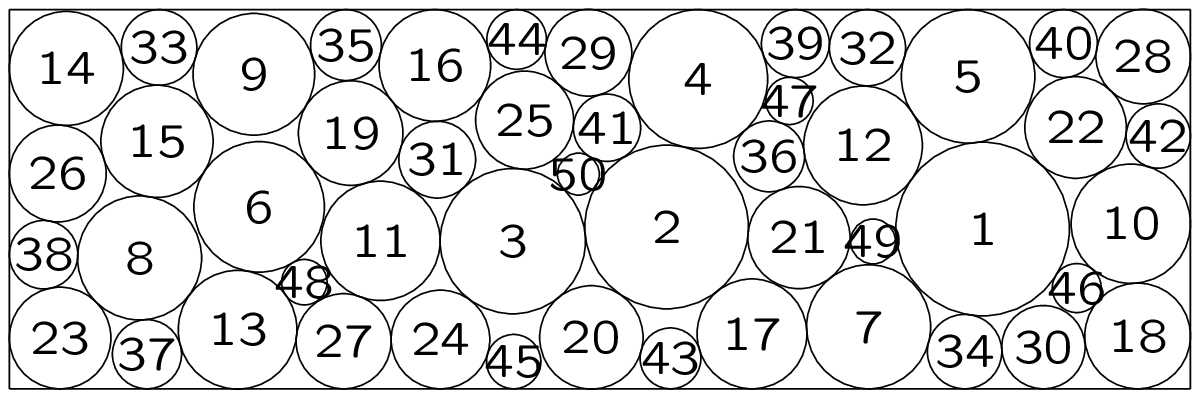}\\
\includegraphics[width=2.5in]{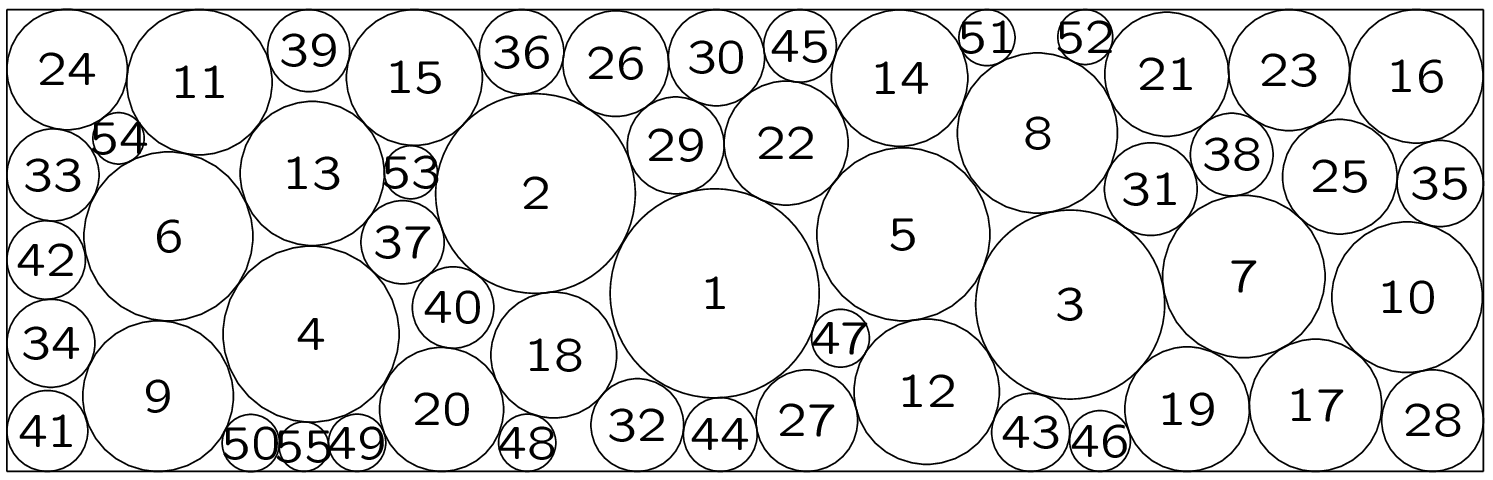}
\includegraphics[width=2.5in]{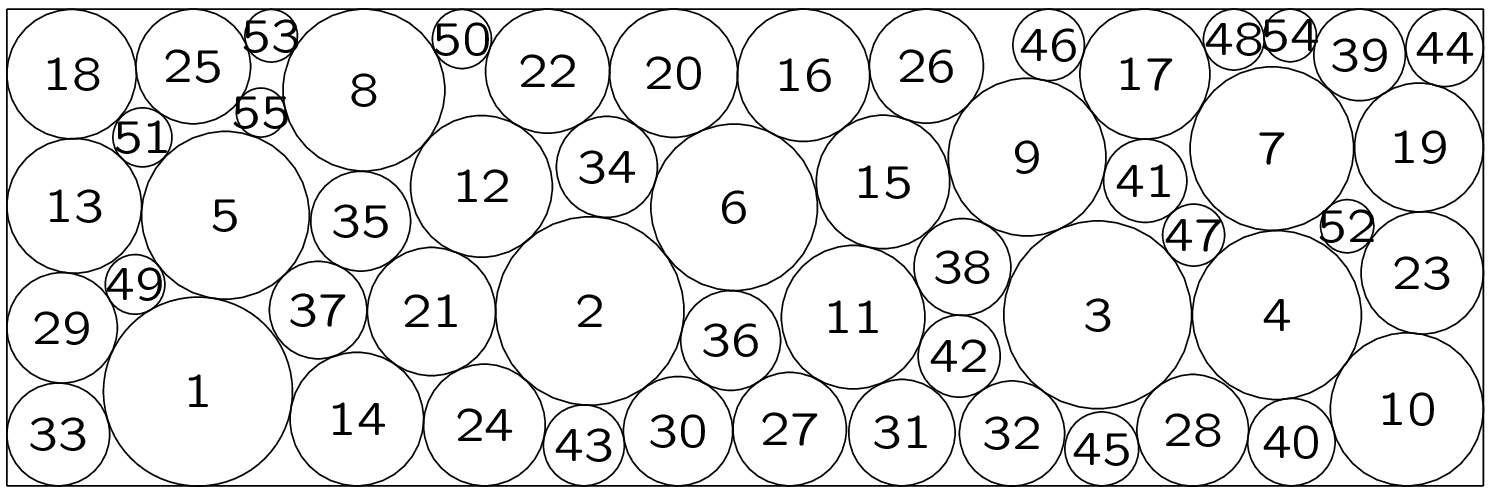}\\
\includegraphics[width=2.5in]{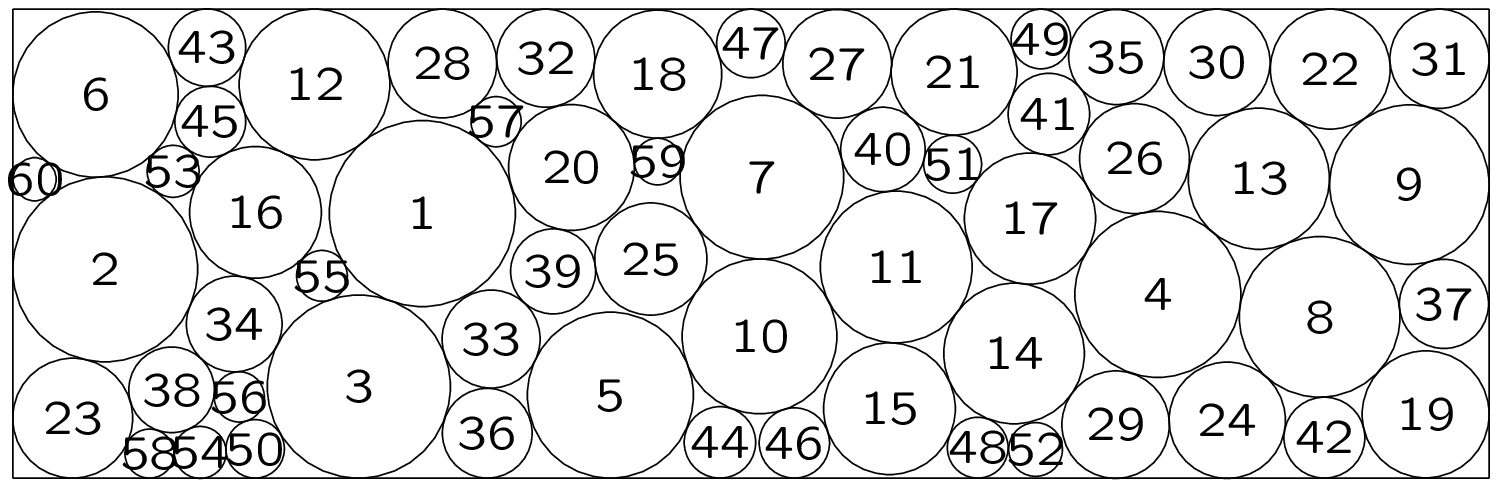}
\includegraphics[width=2.5in]{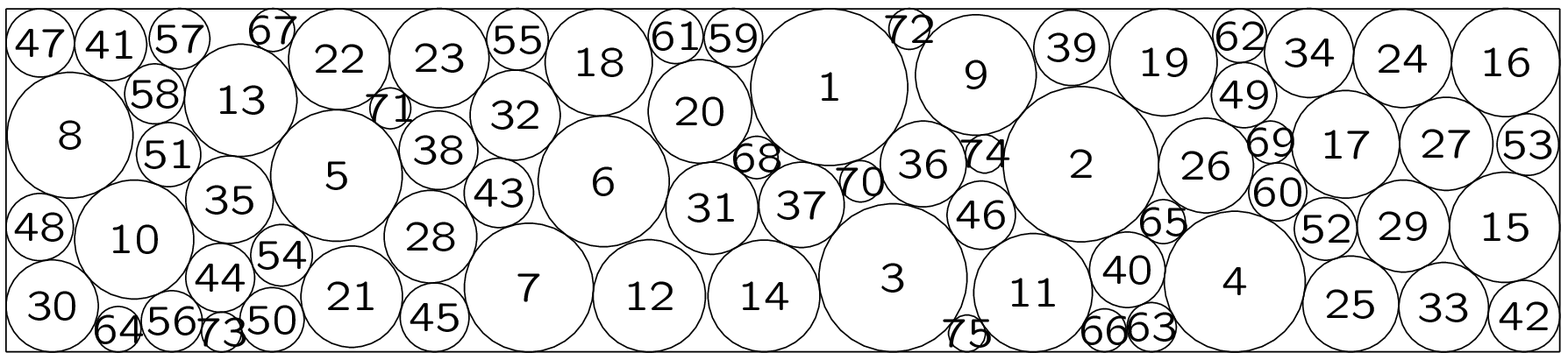}\\
\includegraphics[width=5in]{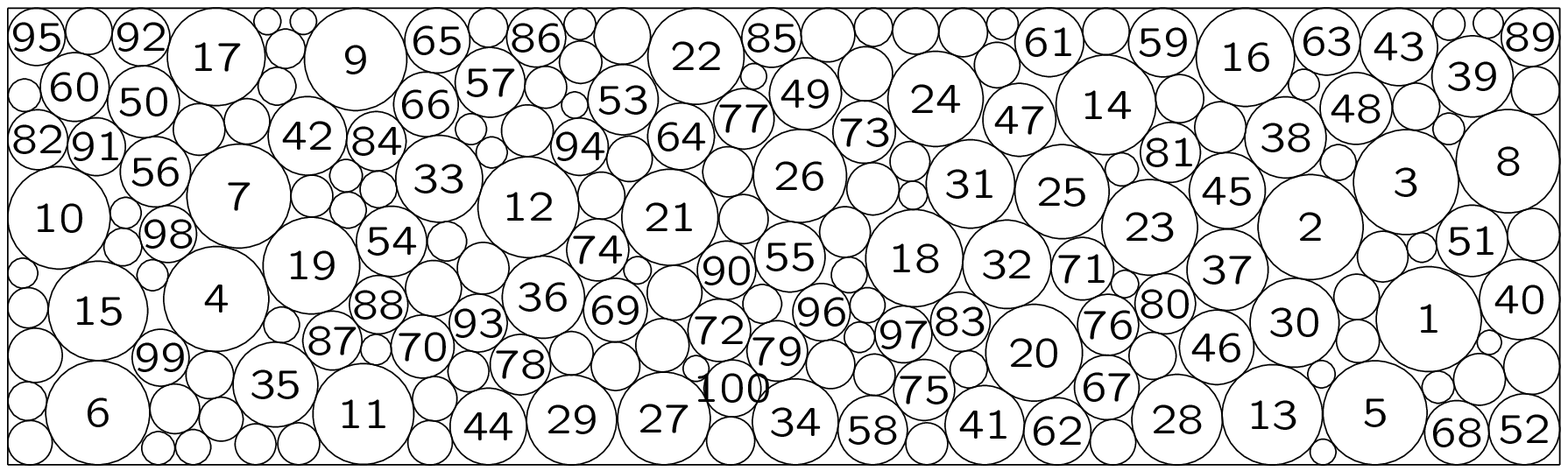}\\
\includegraphics[width=2.5in]{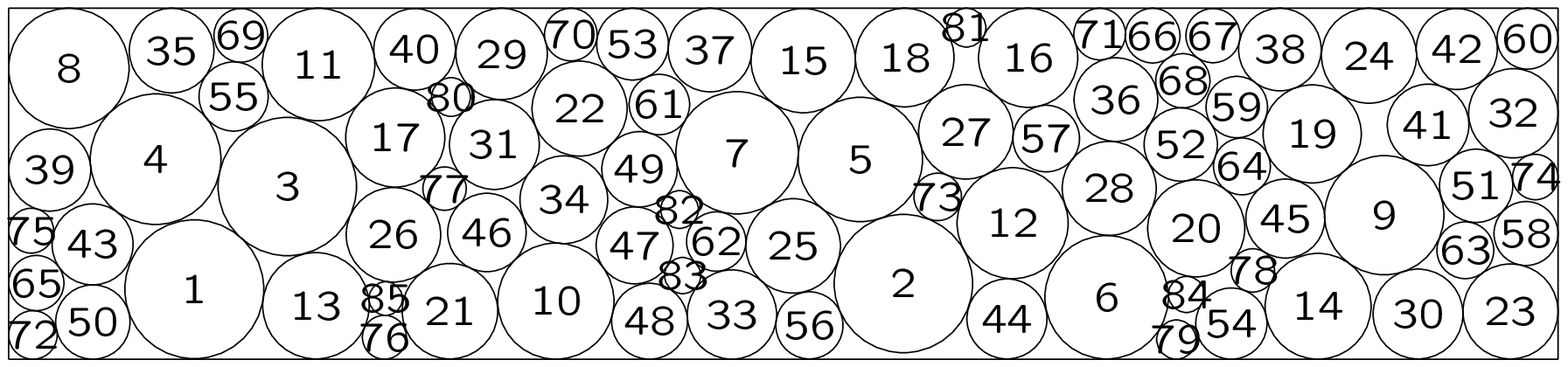}
\includegraphics[width=2.5in]{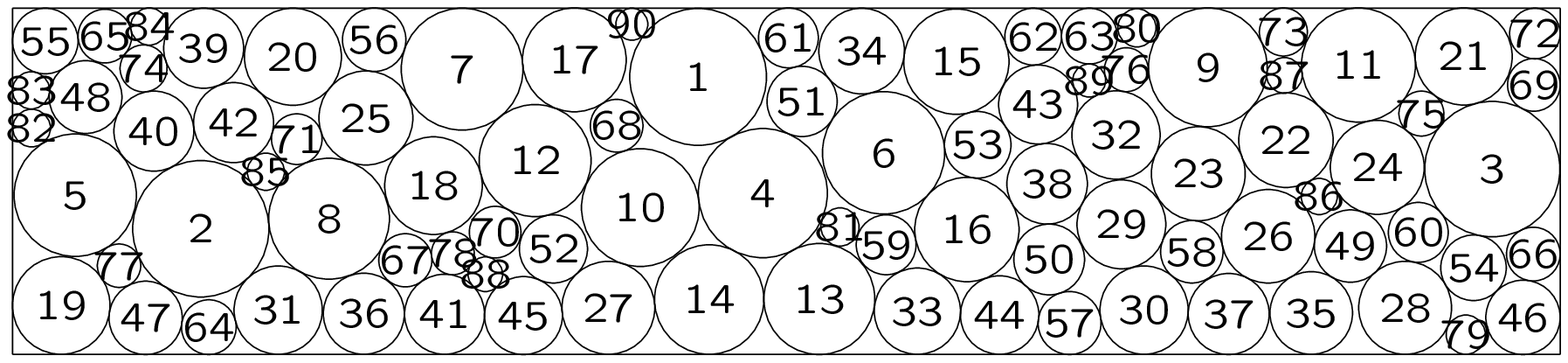}\\
\includegraphics[width=2.5in]{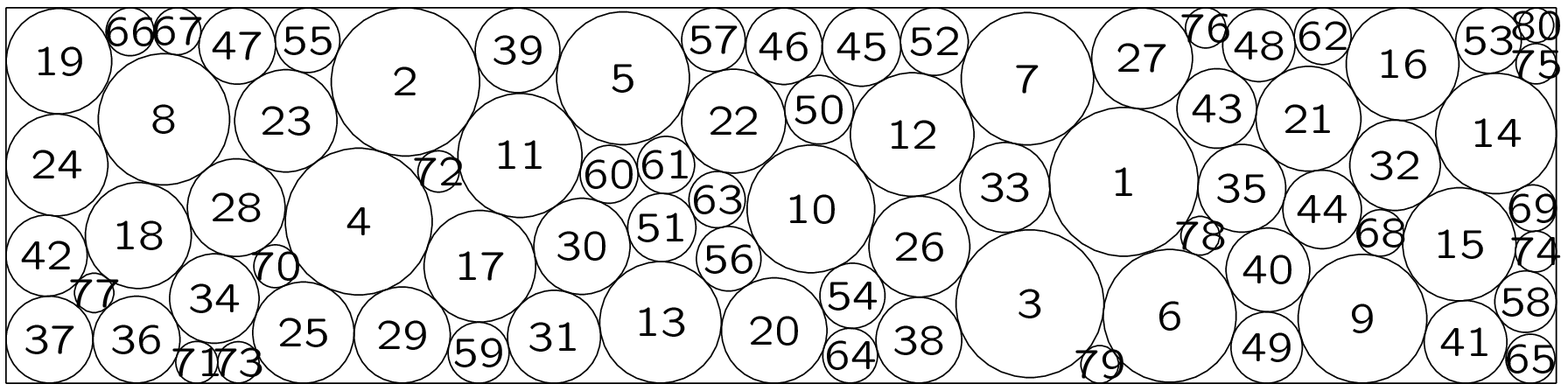}
\includegraphics[width=2.5in]{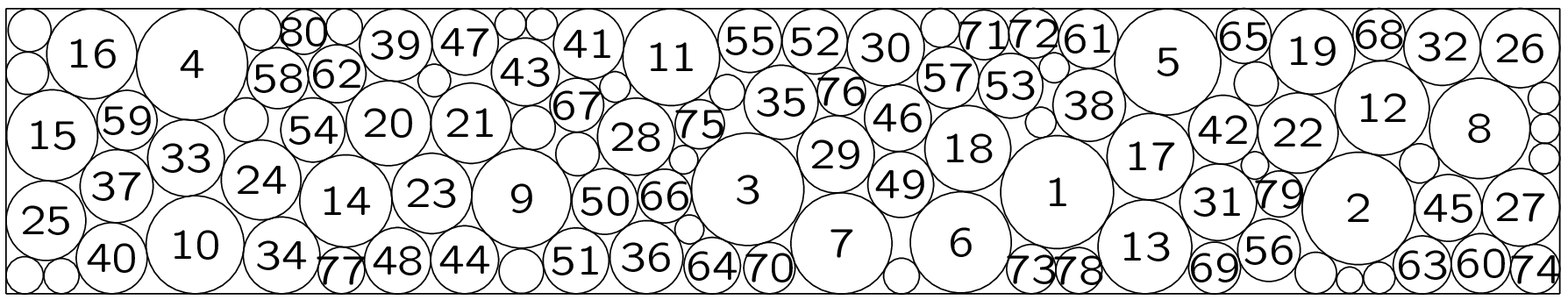}\\

\caption{Improved solutions obtained by ITS for 13 instances (i.e., SY1, SY2, SY3, SY12, SY13, SY24, SY34, SY123, SY56, SY124, SY134, SY234, SY1234 in sequence) }
\end{center}\vspace{-4mm}
\end{figure}

On the other hand, as same as most of the compared approaches, every solution reported by ITS is obtained within 30 hours. Statistically, the average computation time elapsed by ITS for each instance is about 14 hours, while every solution reported by A-SEP-MSBS was obtained after 30 hours, and the average computation time elapsed by MSBS \& SEP-MSBS for each instance was about 24 hours. Note that ITS is run on a computer with 2.87 GHz Processor and 512 MB RAM, while A-SEP-MSBS, MSBS \& SEP-MSBS, BSBIS were all run on a computer with 3 GHz Processor and 256 MB RAM (indeed, less than 10 MB memory is needed to run ITS), it implies that ITS is quiet efficient in term of computation time, with respect to the compared approaches.

In a word, the comparison in terms of both solution quality and computation time demonstrates that ITS is a rather competitive algorithm for solving CODP, with respect to previous state-of-the-art approaches. Furthermore, we would like to point out that for the tested benchmark instances, all the compared algorithms were constructive approaches, which attempted to pack the circles one by one into the container according to some constructive rules. In contrast, ITS is a representative perturbation-based approach, which squeezes all the circles into the container at first, and then attempts to gain further improvements by continuous optimization, tabu search (a special variant of neighborhood search) and solution perturbation. Obviously, these two kinds of approaches are essentially different from each other. Computational experiments have demonstrated the effectiveness of ITS in comparison with several other constructive approaches. We hope the comparison between them would be helpful for further research.

In addition, in order to evaluate the performance of ITS for solving other more general variant of the arbitrary sized circle packing problem (ACP), we have proposed a similar version of the ITS algorithm (with tiny modification of the penalty function, tabu search procedure and perturbation operator) and attempted to resolve the problem of packing arbitrary sized circles into a circular container, so as to minimize the radius of the container. For this variant of ACP, a famous international contest was organized in 2008, with a set of 46 instances ($5\le N\le50$) as benchmarks. For each instance, $r_i=i \ (i=1,\ldots,N)$, respectively. During the competition, 155 groups from 32 countries submitted a total of 27490 solutions online, of which the best results were published in Addis et al. (2008a). Furthermore, M\"uller et al. (2009) proposed a simulated annealing approach and succeeded in improving 26 best known results with $N\ge24$. Based on these 46 instances, Specht extended them to $N=100$ and made all the best known results available on www.packomania.com, which has been a well-known database for different variants of the circle packing problem (CPP).

In our opinions, this set of instances are really representative and extremely challenging, therefore, we use them as benchmarks to evaluate the performance of ITS for dealing with the situation of circular container. Experiments results based on the same hardware platform mentioned above show that ITS succeeds in improving 5 ($N=30, 32, 36, 37, 41$) best known results with $N\le50$, meanwhile, it improves many best known results with $50< N\le100$ (the detailed coordinates are available on www.packomania.com). Note that every result is obtained within 30 hours, while almost all the other researchers did not report the computation time. To summarize, the experiments results demonstrate that ITS is not only suitable for the situation of strip or rectangular container, but also suitable for the situation of circular container, it is an effective approach for solving ACP.

\begin{table}
\noindent {\small Table 2\quad Comparison between ITS (with perturbation operator) and the duplicated TS algorithm (without perturbation operator) }\vspace{-6mm}\\
{\footnotesize
\begin{center} \doublerulesep 0.4pt \tabcolsep 4pt
\begin{tabular}{@{}cccccccc@{}} \hline \hline
\multicolumn{1}{c}{} &\multicolumn{1}{c}{} &\multicolumn{3}{c}{ITS (with perturbation)} &\multicolumn{3}{c}{duplicated TS (without perturbation)} \\ \cmidrule(r){3-5} \cmidrule(r){6-8}
Instance & $L^{best}$  & $L^{ITS}$  &$TS$ & $t_{total}$ &$L^{TS}$ &$TS$ &$t_{total}$  \\ \hline
$SY1$     &17.0954 &\textbf{17.0782}      &2939           &15115s          &17.1252           &19425         &30h        \\
$SY2$     &14.4548 &14.4541               &4674           &8538s           &\textbf{14.4507}  &34428         &52334s     \\
$SY3$     &14.4017 &\textbf{14.3864}      &2300           &8842s           &14.4353           &49357          &30h       \\
$SY4$     &\textbf{23.3538} &23.4177      &13327          &30h             &23.6946           &11612         &30h        \\
$SY5$     &\textbf{35.8590} &35.9843      &359            &30h             &36.1064           &215           &30h        \\
$SY6$     &\textbf{36.4520} &36.6515      &289            &30h             &36.7351           &238           &30h        \\
$SY12$    &29.6837 &\textbf{29.5835}      &377            &18781s          &29.8522           &2237          &30h        \\
$SY13$    &30.3705 &\textbf{30.3621}      &1589           &79168s          &30.8527           &1999          &30h        \\
$SY14$    &\textbf{37.7244} &37.7512      &1483           &30h             &38.3340           &1182          &30h        \\
$SY23$    &\textbf{27.6351} &27.6830      &4672           &30h             &28.0332           &3917          &30h        \\
$SY24$    &34.1455 &\textbf{34.0701}      &726            &29576s          &34.3884           &1901          &30h        \\
$SY34$    &34.6376 &\textbf{34.6263}      &438            &15243s          &35.5036           &1612          &30h        \\
$SY56$    &64.7246 &\textbf{64.5216}      &9              &49126s          &65.0716           &17            &30h        \\
$SY123$   &43.0930 &\textbf{42.9566}      &187            &17684s          &43.5016           &665           &30h        \\
$SY124$   &48.6101 &\textbf{48.5622}      &238            &46715s          &49.0600           &400           &30h        \\
$SY134$   &49.2739 &\textbf{49.2224}      &28             &5259s           &49.6101           &321           &30h        \\
$SY234$   &45.4586 &\textbf{45.4155}      &395            &33586s          &46.1748           &546           &30h        \\
$SY1234$  &60.3346 &\textbf{59.9709}      &63             &29595s          &60.6436           &156           &30h        \\
$Average$ &35.9616 &\textbf{35.9277}      &1894           &$\approx$14h    & 36.3096          &7235          &$\approx$29h  \\

\hline \hline
\end{tabular}
\end{center}} \vspace{2mm}
\end{table}

Finally,in order to analyze the behavior of the perturbation operator, as well as the acceptance criterion, we keep all the other parameters unchanged and then compare the performance of ITS (with perturbation operator) with the duplicated TS algorithm (without perturbation operator). Specifically, the duplicated TS algorithm is obtained by repeatedly launching the tabu search procedure detailed in Algorithm 2, every time from a randomly generated initial solution, until a solution better than the best known solution has been found or the limited 30 hours has been elapsed. It is easy to find out that, if the perturbation operator of ITS is designed to restart randomly every time, ITS becomes the same as the duplicated TS algorithm. Therefore, we hope the comparison between them can reflect the impact of the perturbation operator associated with the acceptance criterion.

Table 2 reports the experimental results. Respectively, column 1 indicates the instance name and column 2 indicates the smallest strip length $L^{best}$ reported by previous approaches. Columns 3-5 indicate the results corresponding with the proposed ITS algorithm, including the strip length $L^{ITS}$ (column 3), the times the tabu search (TS) procedure is re-launched (column 4) and the elapsed time (column 5). For comparison, columns 6-8 indicate the results corresponding with the duplicated TS algorithm, including the value of strip length $L^{TS}$ (column 6), the times the tabu search (TS) procedure is re-launched (column 7), as well as the elapsed time (column 8). Specifically, the results in bold indicate the best ones of all the results reported by various approaches, and \emph{30h} denotes that the limited 30 hours is elapsed for some instances.

As shown in Table 2, ITS succeeds in improving 13 best known results and fails to match the left 5 best known results. In contrast, the duplicated TS algorithm only succeeds in improving the best known result of $SY2$ with smallest size ($N=20$), it fails to match any one of the other best known results. Overall, the average strip length obtained by the duplicated TS algorithm is 36.3096, which is about 1.06\% larger than the average length obtained by ITS (35.9277), and about 0.97\% larger then the average of the best known results (35.9616), respectively. It demonstrates that ITS is rather effective in comparison with the duplicated TS algorithm, especially for instances of large size. In our opinions, it is because that the tabu search procedure only realizes the intensification search within a limited search region, the lack of diversification mechanism weakens the global search capability of the duplicated TS algorithm, especially for large-sized instances.

On the other hand, for each instance, ITS re-launches the tabu search procedure 1894 times on average, the average elapsed time is about 14 hours. In contrast, the duplicated TS algorithm re-launches the tabu search procedure 7235 times for each instance and the average elapsed time is about 29 hours. Apparently, ITS is much more efficient than the duplicated TS algorithm.

 In a word, the comparison in terms of both solution quality and computation efficiency implies that ITS undoubtedly dominates the duplicated TS algorithm, it demonstrates the significance of the perturbation operator, as well as the acceptance criterion.

\section{Conclusion}
Cutting and packing (C\&P) problems are well known NP-hard problems and are widely encountered in practical applications. This paper mainly investigates the circular open dimension problem (CODP), which is a representative variant of the C\&P family. For this problem, an iterated tabu search algorithm named ITS is proposed, which is composed of a tabu search procedure (TS) and a solution perturbation operator associated with an acceptance criterion. As a representative perturbation-based approach, the framework of ITS is quite different from the previously proposed approaches for solving CODP, most of which were constructive approaches. Computational experiments show that ITS produces quite competitive results compared with other state-of-the-art approaches. For two sets of representative problem instances taken from the literature, ITS succeeds in improving the best known results on 13 occasions out of all the 18 instances, the computation time remains reasonable for each instance. 

In addition, supplementary experiments show that ITS is also very effective for solving another closely related variant of CODP: the problem of packing arbitrary sized circles into a circular container. The significance of the perturbation operator as well as the acceptance criterion is also analyzed.

We would like to continue our investigation in the following ways: (1) Further improve the tabu search procedure. (2) Develop some more effective strategies for global perturbation. (3) Consider the situations of packing arbitrary sized circles into a circular, square or triangular container. (4) Extend our investigation to three-dimensional situations. We hope that these attempts would achieve further improvements with the research about C\&P problems.

\section*{Acknowledgements}
This work was supported by National Natural Science Foundation of China (Grant No. 61173180, 61100144 and 61100076). The authors would like to give sincere thanks to Qinghua Wu for his helpful comments and suggestions which improved the quality of this paper.

\bibliographystyle{elsarticle-harv}
\bibliography{<your-bib-database>}







\end{document}